# Understanding and Coping with Hardware and Software Failures in a Very Large Trigger Farm

Jim Kowalkowski
*Fermi National Accelerator Laboratory, Batavia, IL, 60510*

When thousands of processors are involved in performing event filtering on a trigger farm, there is likely to be a large number of failures within the software and hardware systems. BTeV, a proton/antiproton collider experiment at Fermi National Accelerator Laboratory, has designed a trigger, which includes several thousand processors. If fault conditions are not given proper treatment, it is conceivable that this trigger system will experience failures at a high enough rate to have a negative impact on its effectiveness. The RTES (Real Time Embedded Systems) collaboration is a group of physicists, engineers, and computer scientists working to address the problem of reliability in large-scale clusters with real-time constraints such as this. Resulting infrastructure must be highly scalable, verifiable, extensible by users, and dynamically changeable.

## 1. INTRODUCTION

Time critical event filtering applications running on trigger farms with thousands of processors are likely to suffer from a large number of failures within the software and hardware systems. The BTeV experiment [1] includes a trigger with approximately 5,000 CPUs. If fault conditions are not given proper treatment, it is conceivable that this trigger system will experience failures at a high enough rate to have a negative impact on its effectiveness. It is likely that an administrative staff and cast of experiment operators will not be able to service simple problems or analyze complex problems or relationships in a timely fashion to avoid data loss. The Real Time Embedded System collaboration (RTES) [2] is a group of physicists, engineers, and computer scientists, largely funded by an NSF ITR grant [3]. They are working to address the problem of reliability in large-scale clusters with realtime constraints such as this. RTES is defining software infrastructure to detect, diagnose, and recover from errors not only at the system administrative level, but also at the application level. This infrastructure must be highly scalable to minimize bottlenecks or single points of failure. It has to be verifiable to make sure that it does what it is supposed to do in a timely fashion, extensible by users to acquire new detection/analysis methods as they are discovered, and dynamically changeable so that it can be reconfigured as the system operates. The problem is being approached using a hierarchy of monitoring and control elements, architected such that lower levels have high data rates, short reaction times and a narrow view, and higher levels have aggregated data summaries, longer reaction times, and a more global perspective. The purpose of the paper is to describe the RTES group, its relationship to BTeV, the problems they are addressing in regards to fault handling, the solutions they are working on, and some of the difficulties involved in pulling together dissimilar interests.

### 1.1. BTeV Trigger System

The BTeV trigger system is being used as a model for researching fault behavior and handling. A goal of this trigger is to reliably apply processing to every crossing generated by the BTeV pixel and muon detectors. In order to achieve this goal, the designers have divided the trigger into two farms: level-1 and level-2/3. The BTeV detector will produce a crossing every 132ns. With a budget of 2500 embedded processors at level-1, this means that the system must, on average, process an event every 330us. A queuing hierarchy permits the trigger to operate with no fixed time latency (the decision process is not synchronized on a fixed time schedule). The level-2/3 farm has a budget of about 2500 x86 based PCs. Given the rejection requirements, the average processing time per events are about 13ms for the level-2 decision and about 130ms for the level-3 decision.

### 1.2. The Problem

The BTeV trigger serves as a good model for studying fault handling in a large scale distributed computing environment. BTeV requires the trigger to be highly available, sustain high computational performance, and maintain functional integrity over long periods of time. The trigger must be maintainable and be capable of evolving over time to accommodate new ideas and experiments. Given the large number of connection points and commodity parts used in this trigger, we are expecting component hardware failures to occur frequently. This system will contain a large amount of software (and firmware). The reliability of this system is going to depend greatly on the quality of this software, or how well it is tested, how well it is designed, and how well it handles exceptional conditions. Improving the overall quality of the system is going to require quick and easy ways to identity problems and make necessary corrections. This will be especially critical during detector commissioning; when the software and hardware must interact with real-world data for the first time.

### 1.3. The Goal

In order to satisfy the requirements of this trigger and address the problems associated with it, we need a fault handling subsystem. The goal is to create one such subsystem to be used by all components in the trigger and DAQ. This subsystem must be capable of accurately identifying problems and compensating for them. This includes application related activities such as changing algorithm thresholds and overall system activities such as load shifting. As many recovery procedures as possible must be automated. A simple example is switching to a hot spare level-1 processing board when a working board fails. Operators and system developers must be able to easily incorporate new procedures or policies into the system. The operators must be able to easily select error handling policies. A detailed record of observations and actions must be kept to facilitate reproduction of analysis results and to identify long-term trends.

Creating a single subsystem for handling faults across the DAQ and the trigger can benefit the experiment by lowering new procedure integration costs and reducing the amount of knowledge necessary to operate and maintain the system. A standard set of interfaces and protocols reduces the number of conversions and products that must be developed and maintained. Developing standards for error handling and reporting means that the information produced or exchanged between applications can be easily processed.

## 2. THE RTES COLLABORATION

The RTES group is a collaboration of five institutions, funded by NSF Information Technology Grant ACI-0121658. The collaboration consists of physicists, computer scientists, and electrical engineers from University of Illinois, University of Pittsburgh, University of Syracuse, Vanderbilt, and Fermilab. The group has experts in reliability and fault tolerance, real-time scheduling and load balancing, embedded system development, and system modeling. The purpose of this group is to research methodologies and tools for doing fault handling and analysis in large scale, real-time environments. BTeV provides the physical, concrete problem that can be used to demonstrate and benefit directly from the research. BTeV has a real need for this fault handling research and RTES has a real need for a very large-scale system for testing ideas and setting the scope of the research. Pulling together many areas of expertise will lead to a good overall solution. Each of the individual groups adds a unique perspective and number of ideas for solving the problem.

## 3. TECHNOLOGIES

Each of the universities involved has expertise in some aspect of the problem. In some cases, they already have tool kits that have been used to solve smaller scale problems related to realtime embedded systems and fault management. BTeV has established a prototype architecture for the trigger that is being used as a model for RTES software development. The prototype helps to set subsystem boundaries, give a sense of scale, and identify required interfaces and error conditions. This prototype uses DSPs as the embedded level-1 processors. Figure 1 shows a block diagram of the trigger components. This paper concentrates on one of the farmlets and its components, and on the L2/3 node and its software infrastructure. The farmlet is basically a single event input queue (FPGA) with three to six servers. It also contains a microcontroller that is used for configuration, controls, and monitoring.

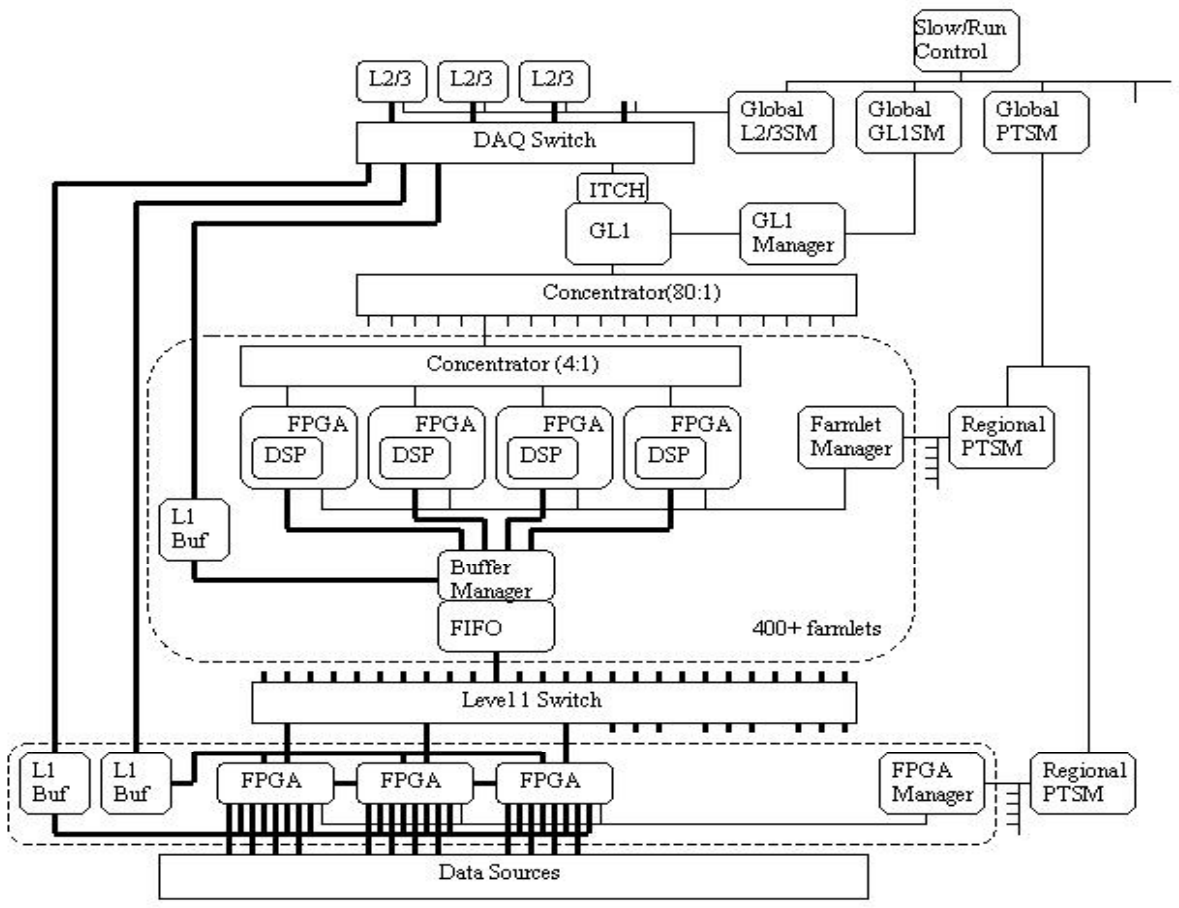

**Figure 1 – Working model of the BTeV trigger**

The technologies introduced by RTES and discussed below are ARMORs for L2/3 nodes and overall management nodes, VLAs for the embedded processors and specific monitoring tasks at L2/3, and GME for system modeling and configuration. Each of these apply to different aspects of the trigger and all of them must work together.

### 3.1. ARMORs

The University of Illinois has produced a fault management software component called an Adaptive, Reconfigurable, and Mobile Objects for Reliability (ARMOR). An implementation of ARMOR exists called Chameleon [4]. ARMORs are multithreaded processes composed of replaceable building blocks called Elements. Elements communicate by way of messages. The pluggable component architecture makes this a highly flexible system allowing modules like recovery action elements, error analysis elements, and problem detection elements to be developed and configured independently. ARMORs can be configured in a hierarchy across multiple nodes to provide entire system coverage. Figure 2 illustrates a simple armor configuration. Here a node has a main ARMOR daemon watching over the node and reporting to higher-level ARMORs out on the network. Elements within these node-level ARMORs work together to make sure all nodes are operating properly. Another standard ARMOR is the execution ARMOR. This ARMOR is responsible for protecting a single application. This type of protection does not require modifications to the program; it simply watches that the program is running. It can restart the application, generate messages for other elements to analyze, or trigger recovery actions based on returns codes from the application.

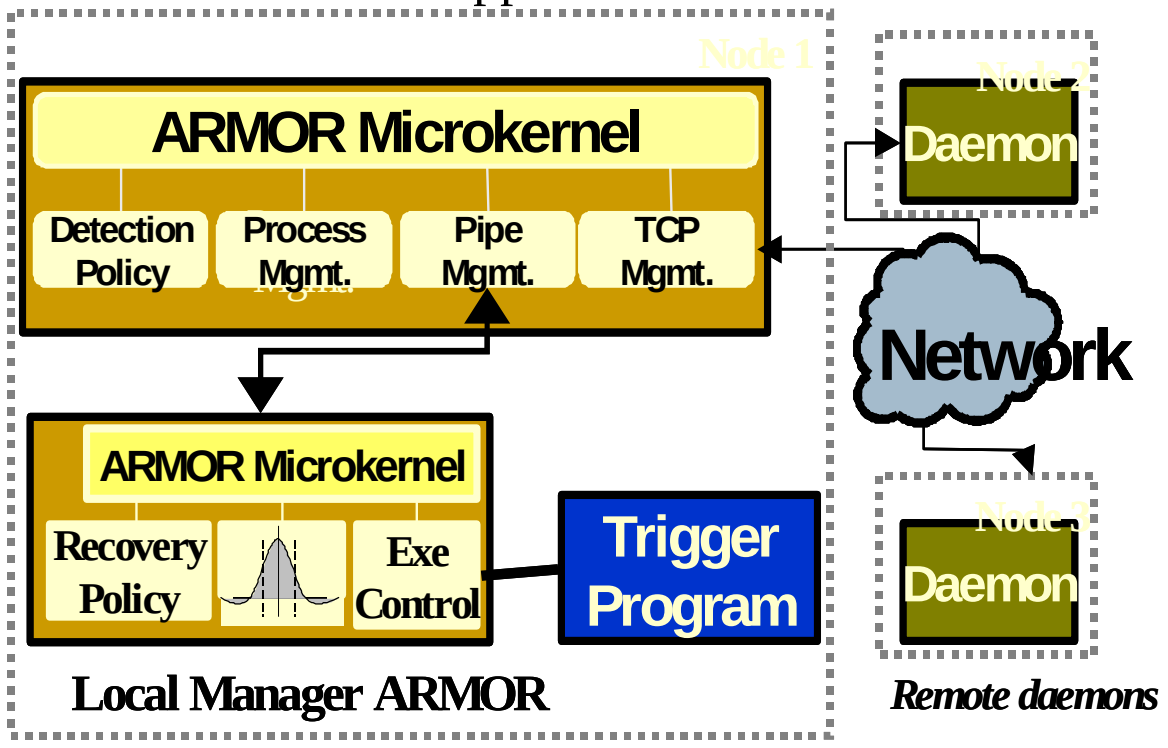

**Figure 2**

An example recovery element is one that automatically migrates processes from one machine to another if a machine is overloaded or has a hardware failure. The ARMORs can also restart at a checkpoint.

Within the trigger, ARMORs can provide error detection and recovery services to the trigger application and any other process running on the L2/3 nodes. They can also watch for hardware failures. ARMORs are designed to run under an operating system such as Linux and Windows and are not well suited for embedded systems with harsh memory and processing time requirements. Using the ARMOR API, the trigger application can report specific errors and other information directly to elements. Data processing rate and data quality measurements can be sent directly into the ARMOR to be distributed to the running elements for analysis. The BTeV online group is currently evaluating ARMORs to watch over DAQ and trigger related processes [5].

### 3.2. VLAs

The University of Syracuse and the University of Pittsburgh are developing a concept called Very Lightweight Agents (VLA). A VLA is a software entity designed to collect various environmental and process related measures, analyze them, and perform actions in a highly constrained environment such as the level-1 trigger of BTeV. VLAs may be realized as a standalone process, a thread, or a collection of functions that maintain state within a larger application [6]. Given memory, CPU time, and network bandwidth constraints, a VLA will decide how to organize itself to provide the best possible results. In order to achieve this goal, a VLA may make use of real-time scheduling, priority queuing, and a hierarchical set of rules to guide its decisions.

Within the context of the level-1 trigger, we imagine VLAs watching for fault conditions such as trigger algorithm crashes, link failures, inability of the processor to keep up, and algorithm running too long. Since the level-1 trigger is so restricted, VLAs will rely on a higher-level control system to do any complex analysis and decision-making. It is easy to believe that running any amount of VLA code during the first part of data taking will cause the processor to fall behind. The VLA will need to be smart enough to change its behavior as data taking progresses, to know problem priorities, and to know when the best time is to report to the larger control system. At level-2/3, VLAs may reside directly inside the trigger executable, performing similar function as in level-1. They will also be used to collect other hardware specific information such as CPU temperature readings.

### 3.3. Modeling Tools

The ISIS group at Vanderbilt University has produced a graphical modeling tool called the Generic Modeling Environment (GME) [7]. The tool enables one to do “Model Integrated Computing” (MIC). This tool allows a designer to model many aspects of a system by creating diagrams. The concept is similar to CASE tools in that it does capture component relationships and properties. It is different because it is not tied to a particular modeling paradigm such as UML [8]. The tool allows the designers to create a domain specific set of rules that define the modeling components; you create a modeling paradigm specific to your project. GME allows one to independently capture different aspects of a system using shapes, properties, associations, and constraints specific to the project and then combine them to form a system image [9]. Just as a compiler forms a parse tree from a programming language and then processes the information in the tree to create machine specific assembly code, GME creates a set of data structure representing the information in the models and allows “model interpreters” to generate information about the system. Typical model interpreters are C/C++ code generators and system configuration generators. Examples of aspects are hardware configuration, process dataflow, and fault handling. Hardware configuration includes physical components and their connectivity. Dataflow includes logical connectivity and executable configuration. Fault handling diagrams show system reactions to problems using hierarchical state machines. The look and feel of the GME in many of the examples we have seen is similar to electronic circuit design tools.

## 4. USE IN THE BTEV TRIGGER

From the controls, monitoring, and configuration perspective, the trigger forms a natural hierarchy. Using this hierarchy in an intelligent way is a must for the system to scale properly. Figure 3 shows the components of the trigger and their relationships and approximate multiplicities. The diagram is broken up into three levels; as we move down, we move closer to the machine. Regional and global management nodes will be PCs running Linux. The level-2/3 branch shown has twenty-five regional management machines each operating 100 trigger nodes (2500 total nodes). The level-1 branch shown has six management machines each operating 100 boards, which each containing four embedded processors and a front end CPU. Each level in the diagram represents a region where a particular technology will be deployed. Software components that sit on the border will likely utilize two different technologies. Event processing and filtering happens at the lowest level and therefore has the most restrictions.

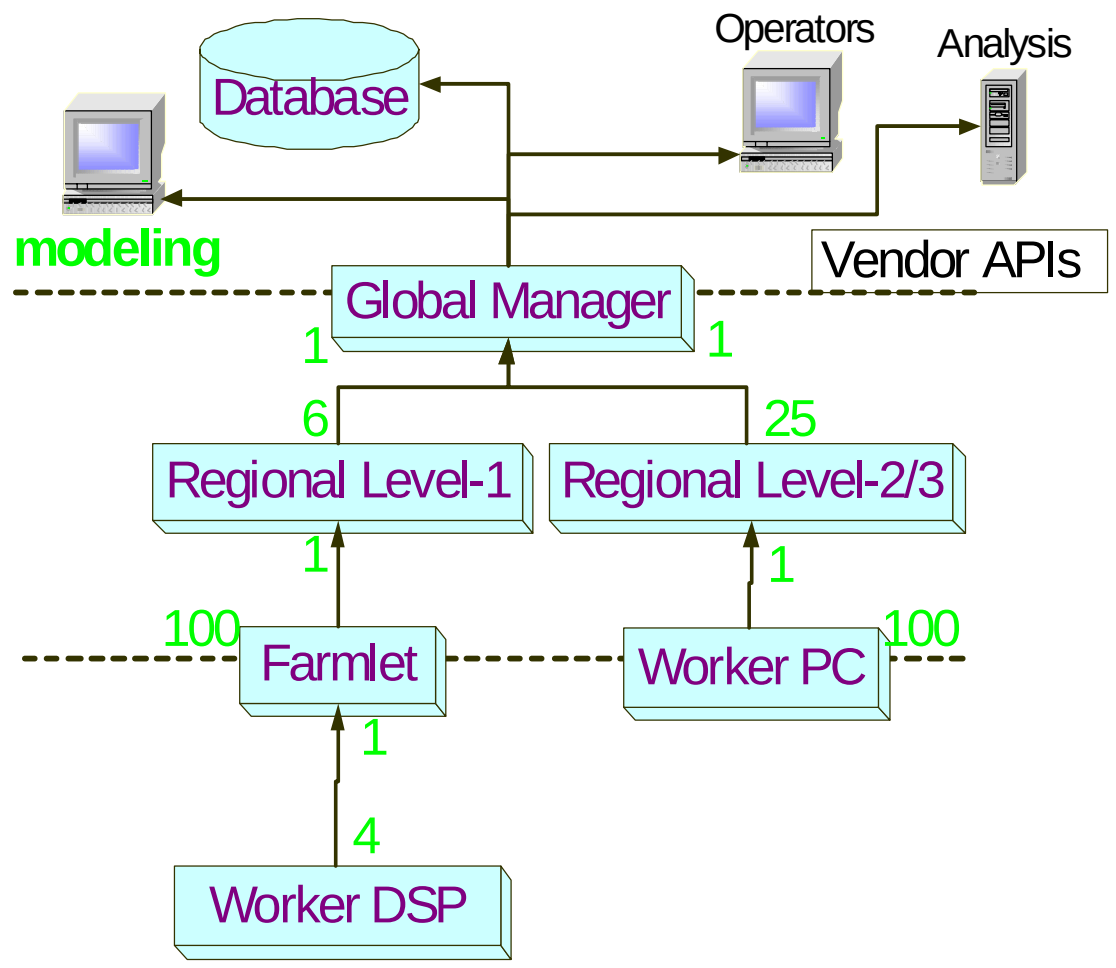

**Figure 3**

The primary technology used on the "Worker DSP", or embedded system level, will be VLAs. Here we will most likely have a simple, lightweight operating system running a single process image containing multiple threads. The operating system will offer basic support for facilities such as resource locking and task preemption and interrupt handling. One plan for the application is for it to be composed of two separately released components: (1) a fault and monitoring subsystem and (2) an event processing subsystem. VLAs will live in subsystem (1). The event processing subsystem can report status and measurements. This subsystem will not necessary know if the measurements are good or bad; it is the job of the fault management subsystem to determine this. VLAs are likely to perform three functions: watch the hardware resources such as communications links, watch for algorithm crashes and timeouts, and watch some of the measurements coming from the trigger algorithm. The VLA infrastructure must be flexible enough to allow dynamic configuration without compilation or application link editing. VLAs may only have time to watch for thresholds or limits to be exceeded and then alert higher-level processes in the Farmlet to do the actual analysis.

ARMORs will handle fault tolerance issues in the "Worker PC" at level-2/3. Here ARMORs will watch over the processes that are active, including the trigger application. The execution elements can restart a failed progress. ARMORs will also manage hardware monitoring VLAs. The VLAs will collect sensor data from the system and summarize it for use in element analysis modules. Examples of system-level VLAs are CPU temperature, network utilization, and I/O error monitoring. VLA code may also be embedded within the physics event building processes and filter program. These VLAs will report event processing statistics and error conditions to the ARMOR for further processing. ARMORs at the "Worker PC" will know how to interpret the data coming in from the various VLAs and know what the proper operating conditions are. They will know how to perform local recovery such as restarting processes and resetting hardware. These ARMORs are tied to a higher-level regional manager. Error conditions that cannot be handled locally will be sent off to this manager for further analysis. Statistical summaries will also be sent up to the manager.

The middle level will likely be ARMOR based. This level has a view across many nodes in the trigger and can perform complex analysis on statistics and errors received from the worker nodes. It can take larger actions such as taking a node out of service or reassigning nodes that are performing other tasks. The top level will handle communications with as many external entities as is needed to operate the trigger. This top-level manager will have the widest view of the system and can perform actions that affect the entire trigger.

## 5. THE ITR EXPERIENCE

### 5.1. Overall Comments

Five institutions make up the RTES group. Each of these collaborating institutions has personal goals they want to achieve and expertise in specific areas of computer science, electrical engineering, and physics. Bringing this diverse set of interests and experiences together forces each group to expand their domain and include new ideas in their thinking. We believe that the overall result will be a better solution to the problem. Working together in this situation also forces everyone to change his or her path toward a solution.

Each group has a set of toolkits and techniques for solving problems they have dealt with in the past. It is natural to first try to apply this existing set to the new problem; to reuse the gained experience and not start from scratch. This can be described as group momentum and it is requires a lot of energy to change course. The scale of the BTeV trigger is larger than any of the groups have seen before and requires changes in process.

The institutions involved are located far apart. Developing a coherent system requires a lot of interaction and communication. The distance between groups impedes progress. It is far more difficult to convene information over the phone or by using email than it is talking person to person. Biweekly teleconferences and bimonthly collaboration meetings help reduce this problem.

We have an overall research goal in mind: creation of methodologies and a toolkit to be used to add fault tolerance to the BTeV trigger. It is not easy to balance individual group research goals and interests with the overall goal. Some research may not be directly useful in a trigger due to constraints, but may be useful in solving a similar large-scale cluster-computing problem. Some of the developments are far different from traditional methods used in the physics community and require an open mind and creative thinking to discover potential uses in a triggering system.

### 5.2. The approach so far

Discussions, prototyping, and simple demonstrations have been the primary tools used so far to move towards the goal. The discussions include concept refinement, such as what a VLA really is, what it means to schedule tasks, what a trigger system does, and how a large-scale system operates (physically and sociologically). RTES sponsored a workshop [10] to discuss ideas and other fault tolerance issues in large systems.

The Vanderbilt group has created DSP boards that mimic the processing that go on in level-1. This platform is being used to understand what types of recovery are possible in this confined environment and how communications take place and are managed. Each of the groups has used this hardware to some extent. Another purpose of this board is to help the group define boundaries between subsystems and develop abstractions and code that are necessary to run actual BTeV level-1 hardware.

### 5.3. Achieving the goal

We are currently in the process of creating a set of requirements that will be used to drive the development process. The interactions between the trigger subsystems and the RTES toolkits will be captured in a set of use cases [11]. The use cases will serve as acceptance criteria for RTES components. A model of the level-1 and level-2/3 trigger system has been defined to set the problem context and aid in the understanding of system boundaries. It also serves as a way for everyone to understand a common set of terms and definitions.

## 6. CONCLUSION

We are working towards having a single fault management toolkit to be used in the BTeV trigger and DAQ systems. This toolkit provides APIs for both system level services and application level programs. It can be used for resource and application monitoring, process management, error reporting, and encapsulation of recovery procedures. The goal is to have a system that is fault tolerant, efficient to operate, and possible to quickly comprehend and extend. This system has an enormous number of resources that must be functional and operating at peak performance to accomplish its task. RTES will enable the operations group to automate problem handling; a must in a system such as this. The expandable nature of RTES components will allow the system to be as smart as the modules that are plugged into it. Careful planning and configuration can lead to an increase in trigger uptime and resource utilization. It will also reduce the time needed to diagnose complex problems and understand the operating characteristics of the software.

The BTeV trigger is a good model for large-scale fault management research. The real-time constraints of the trigger mean interesting research for computer scientists and engineers regarding scheduling and deadline management. We are confident that the new experiences brought in by the RTES group will have a positive effect on BTeV.